\documentclass[11pt]{article}
\usepackage{amssymb,amsmath,amsfonts}
\usepackage{graphicx}
\usepackage{graphics}
\usepackage{eepic,epsfig}


\begin{document}

\title{Topological defects and scalar field modes in warped geometries}
\author{A. A. Saharian$^1$\thanks{%
Corresponding author, E-mail: saharian@ysu.am }, E. L. Karapetyan$^1$, G. V.
Mirzoyan$^{1,2}$ \vspace{0.3cm} \\
\textit{$^1$Institute of Physics, Yerevan State University, }\\
\textit{1 Alex Manoogian Street, 0025 Yerevan, Armenia } \vspace{0.3cm}\\
\textit{$^2$ CANDLE Synchrotron Research Institute, } \\
\textit{31 Acharyan Street, 0040 Yerevan, Armenia}}
\maketitle

\begin{abstract}
We develop a general framework for investigating the influence of
topological defects on the local characteristics of a quantum scalar field
in a warped geometry background. The Ricci tensor and curvature scalar are
decomposed into contributions from the warp factor, the radial geometry and
the angular defect structure. For an arbitrary curvature coupling parameter,
the field equation is separated into independent radial, angular and
warp-coordinate parts. A complete set of normalized mode functions is
obtained for general values of the angular deficit parameters. The general
formalism is applied to several specific cases, such as conformally flat
warped spacetimes, generalized cosmic strings, global monopoles and anti-de
Sitter (AdS)-type warped geometries. The Hadamard two-point function is then
evaluated for a global monopole in AdS spacetime using the obtained mode
functions.
\end{abstract}

\bigskip


\bigskip

\section{Introduction}

Warped geometries constitute an important class of spacetime backgrounds in
contemporary gravitational physics and quantum field theory. They naturally
arise in a wide variety of contexts, including braneworld models,
higher-dimensional cosmology, Kaluza-Klein theories, and effective
descriptions of systems with intrinsic anisotropy or spatial inhomogeneity
(see, e.g., \cite{Gidd06,Lust22,Gied22} and references therein). In such
geometries, the metric structure incorporates a nontrivial warp factor that
modifies the local scale of spacetime, leading to significant curvature
effects even in the absence of conventional matter sources. Important
examples of warped geometries are braneworld models with large extra
dimensions in anti-de Sitter (AdS) background (see reviews in \cite%
{Brax04,Maar10}). In these models, our universe is realized as a
hypersurface (visible brane) in a higher-dimensional spacetime. The
characteristic energy scale of the fundamental gravity is close to the
electroweak energy scale, and the weakness of gravity on the visible brane
is a geometrical effect of the projection. The beaneworld models have been
actively studied as an alternative framework for addressing problems in
particle physics, gravity, and cosmology.

The warped spacetimes provide an interesting ground for studying the
interplay between geometry and quantum phenomena. Quantum fields propagating
on warped backgrounds exhibit behavior that differs qualitatively from their
flat-space counterparts. The presence of curvature affects mode spectra,
vacuum states, and correlation functions, giving rise to phenomena such as
vacuum polarization, particle creation, and induced currents. These effects
have been extensively investigated in time-dependent and cosmological
settings, but geometries with a general structure of the warp factor remain
comparatively less explored, despite their relevance to both fundamental
theory and phenomenological applications. This type of research has
primarily been conducted in the context of anti-de Sitter spacetime,
motivated by applications in braneworld models (see, for example, references
in \cite{Bell22AdS}). In this work, we focus on a class of
higher-dimensional spacetimes characterized by a spatial warp factor
depending on an extra coordinate, combined with a nontrivial internal
geometry of the spatial sections. The geometry under consideration
generalizes standard product manifolds by allowing both warping along an
additional dimension and deviations from exact spherical symmetry in the
angular sector. Such generalized angular structures are motivated by models
with anisotropic compactifications, topological defects, or effective
descriptions of media with direction-dependent properties.

The spacetime investigated here can be viewed as a spatially warped analogue
of time-dependent backgrounds studied previously in \cite{Saha26}. In
contrast to cosmological warp factors, which typically depend on time, the
warp factor in the present setup depends on a spatial coordinate. This
change leads to a qualitatively different curvature structure and alters the
way in which the geometry influences the dynamics of quantum fields. A
central goal of this paper is to provide a systematic geometric analysis of
this class of warped spacetimes in arbitrary dimension. In particular, we
aim to obtain explicit expressions for the Ricci tensor and the scalar
curvature, highlighting how the curvature naturally decomposes into
contributions associated with the warp factor and with the intrinsic
geometry of the angular sector. This decomposition reflects the underlying
product-like structure of the spacetime and proves to be especially useful
for formulating and analyzing field equations transparently.

As an application of the geometric framework, we study the dynamics of a
massive scalar field with an arbitrary curvature coupling. Scalar fields
with nonminimal coupling arise naturally in quantum field theory in curved
spacetime, effective field theories, and models of modified gravity. The
explicit form of the field equation in the present background allows for a
clear separation between contributions stemming from the warp factor, the
radial geometry, and the angular structure. This separation enables a
systematic mode decomposition and facilitates the analysis of spectral
properties. Several physically relevant cases are considered, for which
exact solutions are obtained in terms of special functions. The results
presented in this paper are intended to serve as a foundation for future
studies of quantum vacuum phenomena in spatially warped spacetimes. Possible
applications include the analysis of vacuum expectation values, Casimir-type
effects, and quantum currents, as well as extensions to fields with higher
spin or to interacting theories.

The paper is organized as follows. Section \ref{sec:Geom} describes the
geometric structure of the spacetime and introduces the generalized angular
metric. Section \ref{sec:Modes} is devoted to the mode functions for a
scalar field with a general curvature coupling parameter in a warped
geometry containing topological defects of a general structure. The explicit
expression for the angular part of the mode functions is provided. Section %
\ref{sec:Special} discusses special cases of the general background,
including generalized cosmic strings and global monopoles. As an application
of the mode functions, the Hadamard function for a scalar field is evaluated
in Section \ref{sec:MonAdS} for the geometry of global monopole in
background of anti-de Sitter (AdS) spacetime. Finally, Section \ref{sec:Conc}
summarizes the main results of the paper.

\section{Geometry}

\label{sec:Geom}

The spacetime geometry considered in this work belongs to a class of
higher-dimensional warped manifolds in which the overall structure is
determined by the interplay between an extra spatial dimension and a
nontrivial internal geometry of the spatial sections. A distinctive feature
of the present setup is that the warp factor depends on a spatial coordinate
rather than on time, which leads to a curvature structure qualitatively
different from that of cosmological backgrounds considered in \cite{Saha26}.
In addition, the spatial sections are endowed with a generalized angular
geometry that allows for anisotropic deformations of the standard spherical
metric, thereby extending the class of admissible manifolds beyond maximally
symmetric cases. This generalized angular sector contributes nontrivially to
the curvature and couples to the warp-dependent part of the geometry. As a
consequence, both the Ricci tensor and the scalar curvature acquire
contributions that can be naturally separated into terms associated with the
warping and terms encoding the intrinsic curvature of the angular manifold.
This decomposition plays a central role in the subsequent analysis, as it
clarifies how different geometric features of the spacetime influence the
dynamics of fields propagating on it and provides a transparent framework
for deriving the field equations in arbitrary dimensions.

We consider a $(D+1)$-dimensional warped spacetime, with $D\geq 3$,
described by the line element

\begin{equation}
ds^{2}=dy^{2}+a^{2}\left( y\right) \left[ -dt^{2}+dr^{2}+p^{2}\left(
r\right) d\Omega _{D-2}^{2}\right] ,  \label{ds2}
\end{equation}%
where the warp factor $a(y)$ depends on an extra spatial coordinate $y$,
while the function $p(r)$ determines the intrinsic geometry of the
radial-angular subspace. The angular sector is described by a generalized
metric

\begin{equation}
d\Omega _{D-2}^{2}=\sum_{i,k=1}^{D-2}\gamma _{ik}d\theta _{i}d\theta
_{k}=\alpha _{1}^{2}d\theta _{1}^{2}+\sum_{i=2}^{D-2}\alpha _{i}^{2}\left(
\prod_{l=1}^{i-1}\sin ^{2}\theta _{l}\right) d\theta _{i}^{2},  \label{dOm}
\end{equation}%
which allows for anisotropic deformations of the standard $(D-2)$%
-dimensional sphere through a set of constant parameters $\alpha _{i}$, $%
i=1,2,\ldots ,D-2$. For $D=3$, we have a single angular coordinate $\theta
_{1}$, with the variation range $0\leq \theta _{1}\leq 2\pi $, and the line
element (\ref{ds2}) describes a cosmic string type topological defect with
angle deficit $2\pi (1-\alpha _{1})$. In spatial dimensions $D>3$, the
angular coordinates change in the ranges $0\leq \theta _{i}\leq \pi $ for $%
i=1,\ldots ,D-3$, and $0\leq \theta _{D-2}\leq 2\pi $. For $\alpha _{i}=1$,
the angular part reduces to the usual metric of the unit sphere, whereas
deviations from unity encode anisotropy in the angular directions. The
parameters $\alpha _{i}$ describe the influence of a generalized topological
defect on the background geometry at large distances from the defect's core
(for topological defects and their formation mechanisms see \cite%
{Vile94,Durr02}). The special cases, most frequently considered in the
literature, correspond to cosmic string type defects with $\alpha _{i}=1$
for $i=1,2,\ldots ,D-3$, and $\alpha _{D-2}\neq 1$, and global monopoles
with $\alpha _{1}=\alpha _{2}=\cdots =\alpha _{D-2}\neq 1$.

Due to the warped product structure of the metric, the calculation of the
curvature quantities can be considerably simplified by separating the
contributions that involve the angular sector from those that do not. In
particular, the scalar curvature is conveniently decomposed into two
distinct parts. The first contribution arises from the warping along the
extra spatial coordinate and from the radial-temporal sector of the
geometry, and it contains no dependence on the angular metric. The second
contribution originates entirely from the intrinsic curvature of the angular
subspace and involves the generalized angular metric explicitly. This
separation is introduced purely for technical convenience and does not rely
on any approximation. The angular part of the curvature coincides in form
with the corresponding contribution obtained in previously studied
time-dependent backgrounds and therefore does not require a separate
derivation here. As a result, the calculation can be focused on the
non-angular contribution, which captures all novel features associated with
the spatially dependent warp factor. This decomposition allows one to
isolate the genuinely new geometric effects and greatly facilitates the
subsequent derivation of the field equations.

According to the symmetry of the geometry, the Ricci tensor is diagonal, and
its nonvanishing mixed components naturally split into terms depending on
the warp factor and terms associated with the radial and angular
coordinates. For arbitrary dimension $D$, the nonzero mixed components $%
R_{i}^{i}$, with $i=0,1,D$, do not depend on the angular deficit parameters
and are expressed as%
\begin{equation}
R_{0}^{0}=-\frac{a^{\prime \prime }}{a}-\left( D-1\right) \frac{a^{\prime 2}%
}{a^{2}},\;R_{1}^{1}=R_{0}^{0}-\frac{D-2}{a^{2}p}p^{\prime \prime
},\;R_{D}^{D}=-D\frac{a^{\prime \prime }}{a},  \label{R00}
\end{equation}%
where $a=a\left( y\right) $, $p=p(r)$, $a^{\prime }=da/dy$, and $p^{\prime
}=dp/dr$. The components with the indices in the angular subspace take the
form (no summation over $i$)%
\begin{align}
R_{2}^{2}& =R_{1}^{1}+\frac{D-3}{a^{2}}\left( \frac{p^{\prime \prime }}{p}-%
\frac{p^{\prime 2}}{p^{2}}+\frac{\alpha _{1}^{-2}}{p^{2}}\right) ,  \notag \\
R_{i}^{i}& =R_{i-1}^{i-1}+\frac{(D-1-i)\left( \alpha _{i-1}^{-2}-\alpha
_{i-2}^{-2}\right) }{a^{2}p^{2}\sin ^{2}\theta _{1}\cdots \sin ^{2}\theta
_{i-2}},  \label{Ricci}
\end{align}%
with $i=3,4,...,D-1$. Note that the recurrence relation in (\ref{Ricci}) is
obtained from the corresponding relation for cosmological backgrounds
considered in \cite{Saha26} by the replacement $D\rightarrow D-1$. This is
related to the fact that the spherical part in \cite{Saha26} has a topology $%
S^{D-1}$ whereas in the geometry at hand we consider topology $S^{D-2}$.
From the recurrence relation (\ref{Ricci}) for the components $R_{i}^{i}$
with $i=3,4,...,D-1$, these components are expressed in terms of $R_{2}^{2}$
as%
\begin{equation}
R_{i}^{i}=R_{2}^{2}+\sum_{l=2}^{i-1}\frac{(D-2-l)\left( \alpha
_{l}^{-2}-\alpha _{l-1}^{-2}\right) }{a^{2}p^{2}\sin ^{2}\theta _{1}\cdots
\sin ^{2}\theta _{l-1}}.  \label{Rii}
\end{equation}%
By using the expressions (\ref{Ricci}) and relation (\ref{Rii}), for the
Ricci scalar we get%
\begin{align}
R& =\frac{D-2}{a^{2}}\left[ \left( D-3\right) \frac{\alpha
_{1}^{-2}-p^{\prime 2}}{p^{2}}-\frac{2p^{\prime \prime }}{p}\right]  \notag
\\
& -D\left( D-1\right) \frac{a^{\prime 2}}{a^{2}}-2D\frac{a^{\prime \prime }}{%
a}+\frac{R_{\theta }}{a^{2}p^{2}},  \label{Rscal}
\end{align}%
with the notation%
\begin{equation}
R_{\theta }=\sum_{i=2}^{D-2}(D-i-2)\frac{\left( D-i-1\right) \left( \alpha
_{i}^{-2}-\alpha _{i-1}^{-2}\right) }{\sin ^{2}\theta _{1}\cdots \sin
^{2}\theta _{i-1}},\;D\geq 4.  \label{Rtet}
\end{equation}%
For $D=3$ we have $R_{\theta }=0$. Note that the components of the Ricci
tensor do not depend on $\alpha _{D-2}$.

Let us introduce a new coordinate $z$ in accordance with $dz=dy/a(y)$ and
the function $b(z)=a(y(z))$. In terms of the new coordinate, the line
element is written in the form

\begin{equation}
ds^{2}=b^{2}\left( z\right) \left[ -dt^{2}+dr^{2}+p^{2}(r)d\Omega
_{D-2}^{2}+dz^{2}\right] .  \label{ds2conf}
\end{equation}%
Denoting the derivative with respect to $z$ by $db/dz=\dot{b}$, the
expression for the Ricci scalar takes the form%
\begin{equation}
R=\frac{1}{b^{2}}\left[ R_{z}(z)+R_{r}(r)+\frac{R_{\theta }}{p^{2}}\right] ,
\label{Rscal2}
\end{equation}%
with the notations%
\begin{align}
R_{z}(z) &=-D(D-3)\left( \frac{\dot{b}}{b}\right) ^{2}-2D\frac{\ddot{b}}{b},
\notag \\
R_{r}(r) &=\left( D-2\right) \left[ \left( D-3\right) \left( \frac{\alpha
_{1}^{-2}}{p^{2}}-\frac{p^{\prime 2}}{p^{2}}\right) -\frac{2p^{\prime \prime
}}{p}\right] .  \label{Rr}
\end{align}%
This decomposition isolates the genuinely new geometric effects associated
with the spatially dependent warp factor and provides a transparent
framework for the subsequent analysis of field equations.

\section{Scalar field modes}

\label{sec:Modes}

Here, we consider a massive scalar field $\varphi (x)$ of mass $m$ and
curvature coupling parameter $\xi $, propagating in $(D+1)$-dimensional
spacetime with the line element (\ref{ds2}). The dynamics of the field is
governed by the Klein-Gordon equation

\begin{equation}
\left( g^{ik}\nabla _{i}\nabla _{k}-\xi R-m^{2}\right) \varphi (x)=0,
\label{Feq}
\end{equation}%
where $R$ is the Ricci scalar given by (\ref{Rscal2}). Canonical
quantization of fields begins with the selection of a complete set of
positive and negative energy\ solutions of the classical equations of motion
(\ref{Feq}), denoted here by $\{\varphi _{\sigma }^{(+)}(x),\varphi _{\sigma
}^{(-)}(x)\}$. The collective index $\sigma $ defines the set of quantum
numbers uniquely specifying the mode functions. The latter are normalized by
the condition 
\begin{equation}
\int \ d^{D}x\sqrt{|g|}g^{00}\left[ \varphi _{\sigma ^{\prime }}^{(+)\ast
}(x)\partial _{t}\varphi _{\sigma }^{(+)}(x)-\varphi _{\sigma
}^{(+)}(x)\partial _{t}\varphi _{\sigma ^{\prime }}^{(+)\ast }(x)\right]
=i\delta _{\sigma \sigma ^{\prime }},  \label{norm}
\end{equation}%
where the star corresponds to complex conjugate. The symbol $\delta _{\sigma
\sigma ^{\prime }}$ is understood as Dirac delta function for continuous
components of the collective set $\sigma $ and as the Kronecker delta for
discrete components.The expansion of the field operator, given by%
\begin{equation}
\varphi (x)=\sum_{\sigma }\left[ a_{\sigma }\varphi _{\sigma
}^{(+)}(x)+b_{\sigma }^{\dagger }\varphi _{\sigma }^{(-)}(x)\right] ,
\label{phiexp}
\end{equation}%
defines the annihilation operator $a_{\sigma }$ for particles and the
creation operator $b_{\sigma }^{\dagger }$ for antiparticles. They obey the
standard commutation relations for bosonic fields (for quantization of
fields in curved backgrounds see, e.g., \cite{Birr82,Park11}). In (\ref%
{phiexp}), a summation is understood over discrete quantum numbers and an
integration over continuous ones. The vacuum state $\left\vert
0\right\rangle $ is defined by the properties $a_{\sigma }\left\vert
0\right\rangle =0$ and $b_{\sigma }\left\vert 0\right\rangle =0$. We are
interested in the complete set of mode functions for the field $\varphi (x)$
in background of the geometry described in the previous section.

According to the symmetry of the line element (\ref{ds2conf}), the solution
for the mode functions corresponding to the states with positive energy $E$\
can be presented in the separated form 
\begin{equation}
\varphi _{\sigma }^{(+)}(x)=e^{-iEt}U(r)Z(z)Y(\theta ),  \label{phisep}
\end{equation}%
where we use the collective notation $\theta =(\theta _{1},\theta
_{2},\ldots ,\theta _{D-2})$ for the angular coordinates. Substituting into
the equation (\ref{Feq}) and separating the variables, one obtains
independent equations for the separate factors. The equations for the first
two factors read%
\begin{eqnarray}
\frac{\left[ p^{D-2}U^{\prime }(r)\right] ^{\prime }}{p^{D-2}}+\left[
\lambda _{r}^{2}-\xi R_{r}(r)-\frac{\gamma }{p^{2}}\right] U(r) &=&0,
\label{Ueq} \\
b^{1-D}\frac{d}{dz}\left[ b^{D-1}\dot{Z}(z)\right] +\left[ \lambda
_{z}^{2}-\xi R_{z}(z)-m^{2}b^{2}\right] Z(z) &=&0,  \label{Zeq}
\end{eqnarray}%
where $\lambda _{z}^{2}=E^{2}-\lambda _{r}^{2}$, the prime and dot stand for
derivatives with respect to $r$ and $z$, respectively, and $\gamma $ and $%
\lambda _{r}^{2}$ are the separation constants.

For $D=3$ we have $R_{\theta }=0$ and the corresponding part in the mode
functions is expressed as $Y(\theta _{1})=\mathrm{const}\cdot
e^{in_{1}\theta _{1}}$ with $n_{1}=0,\pm 1,\pm 2,\ldots $. As it has been
mentioned above, this corresponds to a cosmic string type topological
defect. In spatial dimensions $D>3$, for the angular function we get%
\begin{equation}
\left[ \frac{\partial _{\theta _{1}}\left( \sin ^{D-3}\theta _{1}\partial
_{\theta _{1}}\right) }{\alpha _{1}^{2}\sin ^{D-3}\theta _{1}}%
+\sum_{i=2}^{D-2}\frac{\partial _{\theta _{i}}\left( \sin ^{D-i-2}\theta
_{i}\partial _{\theta _{i}}\right) }{\alpha _{i}^{2}\sin ^{D-i-2}\theta
_{i}\prod_{l=1}^{i-1}\sin ^{2}\theta _{l}}-\xi R_{\theta }+\gamma \right]
Y(\theta )=0.  \label{Yeq}
\end{equation}%
The structure of the solution for this equation has been discussed in \cite%
{Saha26}. It is presented in a separated form%
\begin{equation}
Y(\theta )=e^{in_{D-2}\theta _{D-2}}\prod_{l=1}^{D-3}Y_{(l)}(\theta _{l}),
\label{Y1}
\end{equation}%
where $n_{D-2}=0,\pm 1,\pm 2,\ldots $. The equations for separate functions $%
Y_{(l)}(\theta _{l})$, with $l=1,2,\ldots ,D-3$, are obtained from the
equation (\ref{Yeq}). The corresponding solutions, regular at $\theta _{l}=0$%
, are expressed in terms of the associated Legendre function of the first
kind:%
\begin{equation}
Y_{(l)}(\theta _{l})=\mathrm{const}\frac{P_{\nu _{l}}^{-\mu _{l}}(\cos
\theta _{l})}{\sin ^{\frac{D-3-l}{2}}\theta _{l}},\;l=1,\ldots ,D-3,
\label{Yl}
\end{equation}%
where the order and degree are connected by the relation%
\begin{equation}
\mu _{l}^{2}=\frac{\alpha _{l}^{2}}{\alpha _{l+1}^{2}}\left( \nu
_{l+1}+1/2\right) ^{2}+(D-l-3)\left[ (D-l-2)\xi -\frac{D-l-3}{4}\right]
\left( \frac{\alpha _{l}^{2}}{\alpha _{l+1}^{2}}-1\right) .  \label{relmu}
\end{equation}%
Here, 
\begin{align}
\nu _{1} =&\sqrt{(D-3)^{2}/4+\alpha _{1}^{2}\gamma }-1/2,  \notag \\
\nu _{D-2} =&|n_{D-2}|-1/2.  \label{nuD2}
\end{align}%
From (\ref{relmu}) we get%
\begin{equation}
\mu _{D-3}=\frac{\alpha _{D-3}}{\alpha _{D-2}}|n_{D-2}|.  \label{muD3}
\end{equation}%
The regularity condition at $\theta _{l}=\pi $ gives the relation $\nu
_{l}-\mu _{l}=n_{l}$, with $n_{l}=0,\pm 1,\pm 2,\ldots $, between the degree
and order of the associated Legendre function. Combining this relation with (%
\ref{relmu}), we get a recurrence relation between $\nu _{l}$ and $\nu
_{l+1} $. Having $\nu _{D-2}$ from (\ref{nuD2}), it allows to find $\nu _{l}$
for $l=1,\ldots ,D-3$. Then, from (\ref{nuD2}), we can find the separation
constant $\gamma $ entering in the equation (\ref{Ueq}) for the radial
function $U(r)$. Having $\gamma $, the eigenvalues for the quantum numbers $%
\lambda _{r}$ and $\lambda _{z}$ are obtained by solving the equations (\ref%
{Ueq}) and (\ref{Zeq}). Hence, as a complete set of quantum numbers
specifying the modes we can take $\sigma =(\lambda _{r},\lambda _{z},\mathbf{%
n})$ with $\mathbf{n}=\left( n_{1},\ldots ,n_{D-2}\right) $.

The angular part of the mode functions is specified by the set $\mathbf{n}$.
We denote by $Y(\theta )=Y_{\mathbf{n}}(\theta )$ these functions normalized
by the condition%
\begin{equation}
\int d^{D-2}\theta \,\sqrt{\mathrm{det\,}\gamma _{ik}}Y_{\mathbf{n}^{\prime
}}^{\ast }(\theta )Y_{\mathbf{n}}(\theta )=\delta _{\mathbf{nn}^{\prime }},
\label{Ynorm}
\end{equation}%
where the metric tensor in the angular subspace is given by (\ref{dOm}). The
normalized mode functions become%
\begin{equation}
Y_{\mathbf{n}}(\theta )=C_{\theta }e^{in_{D-2}\theta _{D-2}}\prod_{l=1}^{D-3}%
\frac{P_{\nu _{l}}^{n_{l}-\nu _{l}}(\cos \theta _{l})}{\sin ^{\frac{D-3-l}{2}%
}\theta _{l}},  \label{Y2}
\end{equation}%
where the normalization constant $C_{\theta }$ is determined from the
condition (\ref{Ynorm}). Assuming that the radial functions $U(r)=U_{\lambda
_{r}}(r)$ are normalized by the condition 
\begin{equation}
\int dr\,p^{D-2}U_{\lambda _{r}}(r)U_{\lambda _{r}^{\prime }}^{\ast
}(r)=\delta _{\lambda _{r}\lambda _{r}^{\prime }},  \label{Unorm}
\end{equation}%
for the modes $Z(z)=Z_{\lambda _{z}}(z)$ we get the condition%
\begin{equation}
\int \ dz\,b^{D-1}Z_{\lambda _{z}}(z)Z_{\lambda _{z}}^{\ast }(z)=\frac{%
\delta _{\lambda _{z}\lambda _{z}^{\prime }}}{2E}.  \label{Znorm}
\end{equation}%
In (\ref{Unorm}) and (\ref{Znorm}), the integration is performed over the
entire range of variation of the coordinates $r$ and $z$. These variation
ranges depend on the specific geometry fixed by the functions $p(r)$ and $%
b(z)$.

Introducing the functions $\psi _{r}(r)$ and $\psi _{z}(z)$ in accordance
with 
\begin{equation}
U(r)=p^{-\frac{D-2}{2}}\psi _{r}(r),\;Z(z)=b^{-\frac{D-1}{2}}\psi _{z}(z),
\label{psirz}
\end{equation}%
the corresponding equations are written in the form of the Schr\"{o}dinger
equations%
\begin{equation}
\frac{d^{2}\psi _{u}}{du^{2}}+\left[ \lambda _{u}^{2}-V_{u}(u)\right] \psi
_{u}=0,  \label{psiEq}
\end{equation}%
with $u=r,z$ and "potential energies"%
\begin{align}
V_{r}(r) &=\left( \frac{D-3}{\alpha _{1}^{2}}\xi +\frac{\gamma }{D-2}\right) 
\frac{1}{p^{2}}+\left[ \frac{D-4}{4}-\left( D-3\right) \xi \right] \frac{%
p^{\prime 2}}{p^{2}}+\left( \frac{1}{2}-2\xi \right) \frac{p^{\prime \prime }%
}{p},  \notag \\
V_{z}(z) &=b^{2}m^{2}+D\left( \xi _{D}-\xi \right) \left[ (D-3)\frac{\dot{b}%
^{2}}{b^{2}}+2\frac{\ddot{b}}{b}\right] .  \label{Pots}
\end{align}%
Here, $\xi _{D}=(D-1)/(4D)$ is the curvature coupling parameter for a
conformally coupled scalar field.

\section{Special cases of background spacetime}

\label{sec:Special}

In this section we consider some special cases of background spacetime.

\subsection{Conformally flat specetimes}

First, we take a background geometry that is conformally flat in the absence
of topological defects. This corresponds to the choice $p(r)=r$, with $0\leq
r<\infty $. In the absence of defects one has $\alpha _{i}=1$ for $%
i=1,2,\ldots ,D-2$, and we can introduce Cartesian coordinates $%
(X^{1},\ldots ,X^{D-1})$ in accordance with%
\begin{align}
X^{1}& =r\cos \theta _{1},\;X^{i}=r\sin \theta _{1}\cdots \sin \theta
_{i-1}\cos \theta _{i},  \notag \\
X^{D-1}& =r\sin \theta _{1}\cdots \sin \theta _{D-3}\sin \theta _{D-2},
\label{Xcoord}
\end{align}%
where $i=2,\ldots ,D-2$. The line element is written in explicitly
conformally flat form%
\begin{equation}
ds^{2}=b^{2}\left( z\right) \left[ -dt^{2}+\sum_{i=1}^{D-1}\left(
dX^{i}\right) ^{2}+dz^{2}\right] .  \label{ds2flat}
\end{equation}%
This line element is conformally related to the patch of the AdS spacetime
covered by Poincar\'{e} coordinates (see below).

The presence of defects, in general, alters both the local and global
geometries introducing additional terms in the curvature tensor. In the
special case under consideration with $p(r)=r$, the radial part of the Ricci
scalar and the effective potential become%
\begin{align}
R_{r}(r)=& \left( D-2\right) \left( D-3\right) \frac{\alpha _{1}^{-2}-1}{%
r^{2}},  \label{Rrflat} \\
V_{r}(r)=& \left[ \left( D-3\right) \xi \left( \alpha _{1}^{-2}-1\right) +%
\frac{\gamma }{D-2}+\frac{D-4}{4}\right] \frac{1}{r^{2}}.  \label{Vr1}
\end{align}%
The equation (\ref{Ueq}) for the radial function is reduced to%
\begin{equation}
U^{\prime \prime }(r)+\frac{D-2}{r}U^{\prime }(r)+\left[ \lambda
_{r}^{2}-\left( D-2\right) \left( D-3\right) \xi \frac{\alpha _{1}^{-2}-1}{%
r^{2}}-\frac{\gamma }{r^{2}}\right] U(r)=0,  \label{Ueqflat}
\end{equation}%
with $0\leq \lambda _{r}<0$. The solution of this equation, square
integrable at $r=0$, is given by%
\begin{equation}
U(r)=\sqrt{\lambda _{r}}\frac{J_{\nu _{r}}(\lambda _{r}r)}{r^{\frac{D-3}{2}}}%
,\;\nu _{r}=\sqrt{\frac{(D-3)^{2}}{4}+\gamma +\left( D-2\right) \left(
D-3\right) \xi \left( \alpha _{1}^{-2}-1\right) },  \label{Uflat}
\end{equation}%
where $J_{\nu _{r}}(y)$ is the Bessel function. Here, the normalization
coefficient is determined from (\ref{Unorm}), by using the formula%
\begin{equation}
\int_{0}^{\infty }dr\,rJ_{\nu _{r}}(\lambda _{r}r)J_{\nu _{r}}(\lambda
_{r}^{\prime }r)=\frac{\delta (\lambda _{r}^{\prime }-\lambda _{r})}{\lambda
_{r}}.  \label{Jint}
\end{equation}

The choice of the Bessel function in (\ref{Uflat}) corresponds to the
so-called Dirichlet boundary condition at the origin $r=0$. In special cases
of the parameters, the solution of the equation (\ref{Ueqflat}) with the
Neumann function $Y_{\nu _{r}}(\lambda _{r}r)$ can also be square integrable
at $r=0$. The corresponding values of the parameters are determined from the
condition $\nu _{r}<1$. Under this condition, for the function $U(r)$, we
can take the linear combination of the Bessel and Neumann functions. One of
the coefficients in this combination is determined by the normalization
condition, while the value of the second coefficient is fixed by the
boundary condition at the origin. In the special case of a global monopole
(see below) in three spatial dimensions, this type of boundary condition was
discussed in \cite{Pite09,Camp21}.

\subsection{Models with generalized cosmic strings}

For a generalized cosmic string $\alpha _{i}=1$ for $i=1,\ldots ,D-3$ and $%
p(r)=r$. In this case, for points outside the defect core, corresponding to $%
r>0$, the curvature tensor coincides with the corresponding tensor in the
geometry without defects. In particular, $R_{i}^{i}=R_{0}^{0}$ for $%
i=1,\ldots ,D-1$. For $\alpha _{D-2}\equiv \alpha \neq 1$, the curvature
tensor has delta-type singularity at $r=0$.

It is convenient to pass to cylindrical spatial coordinates $X^{i}=\left( 
\mathbf{X},X^{D-2}=\rho ,X^{D-1}=\phi ,X^{D}=z\right) $ with $\mathbf{X}%
=(X^{1},\ldots ,X^{D-3})$. The line element becomes%
\begin{equation}
ds^{2}=b^{2}\left( z\right) \left( -dt^{2}+d\mathbf{X}^{2}+d\rho ^{2}+\alpha
^{2}\rho ^{2}d\phi ^{2}+dz^{2}\right) ,  \label{ds2cs}
\end{equation}%
with the coordinate ranges $-\infty <X^{i}<+\infty $, $i=1,\ldots ,D-3$, $%
0\leq \rho <\infty $, and $0\leq \phi \leq 2\pi $. In these new coordinates,
the components of the Ricci tensor are expressed as (no summation over $i$) 
\begin{equation}
R_{i}^{i}=-\frac{1}{b^{2}}\left[ \left( D-2\right) \left( \frac{\dot{b}}{b}%
\right) ^{2}+\frac{\ddot{b}}{b}\right] ,\;R_{D}^{D}=\frac{D}{b^{2}}\left[
\left( \frac{\dot{b}}{b}\right) ^{2}-\frac{\ddot{b}}{b}\right] ,
\label{Riics}
\end{equation}%
for $i=0,1,\ldots ,D-1$. Here, as before, the overdot means a derivative
with respect to $z$. For the curvature scalar one obtains 
\begin{equation}
R=\frac{R_{z}(z)}{b^{2}},  \label{Rcs}
\end{equation}%
where the function $R_{z}(z)$ is given by (\ref{Rr}). The core of the
defect, described by the line element (\ref{ds2cs}), corresponding to $\rho
=0$, presents a $(D-2)$-dimensional hypersurface covered by the coordinates $%
(\mathbf{X},z)$.

The factorized positive energy mode functions read%
\begin{equation}
\varphi _{\sigma }^{(+)}(x)=\frac{e^{i\mathbf{kX}+in\phi -iEt}}{\left( 2\pi
\right) ^{\frac{D}{2}-1}\alpha ^{\frac{1}{2}}}\Psi (\rho )Z(z),
\label{phics}
\end{equation}%
where $\mathbf{k}=(k^{1},\ldots ,k^{D-3})$, $-\infty <k^{l}<+\infty $, $%
\mathbf{kX}=\sum\nolimits_{l=1}^{D-3}k^{l}X^{l}$, and $n=0,\pm 1,\pm
2,\ldots $. The factor $\left( 2\pi \right) ^{2-D}/\alpha $ comes from the
normalization of the $\mathbf{X}$- and $\phi $-dependent parts in the mode
functions. Separating the variables in the field equation (\ref{Feq}), we
get the equation (\ref{Zeq}) for the function $Z(z)$ and the equation 
\begin{equation}
\left[ \frac{1}{\rho }\partial _{\rho }\left( \rho \partial _{\rho }\right)
+\lambda _{\rho }^{2}-\frac{n^{2}}{\alpha ^{2}\rho ^{2}}\right] \Psi =0,
\label{Eqscs}
\end{equation}%
for the function $\Psi (\rho )$, where $\lambda _{\rho
}^{2}=E^{2}-k^{2}-\lambda _{z}^{2}$ with $k^{2}=\sum_{l=1}^{D-3}(k^{l})^{2}$%
. As a set of quantum numbers specifying the modes, we can take $\sigma =(%
\mathbf{k},\lambda _{\rho },n,\lambda _{z})$. The energy of the mode is
expressed as $E=\sqrt{k^{2}+\lambda _{\rho }^{2}+\lambda _{z}^{2}}$.

The solution for $\Psi (\rho )$, regular at $\rho =0$ and normalized by the
condition 
\begin{equation}
\int_{0}^{\infty }d\rho \,\rho \Psi _{\lambda _{\rho }}(\rho )\Psi _{\lambda
_{\rho }^{\prime }}^{\ast }(\rho )=\delta \left( \lambda _{\rho }-\lambda
_{\rho }^{\prime }\right) ,  \label{Psinorm}
\end{equation}%
with $0\leq \lambda _{\rho }<\infty $, is given is given in terms of the
Bessel function:%
\begin{equation}
\Psi _{\lambda _{\rho }}(\rho )=\sqrt{\lambda _{\rho }}J_{|n|/\alpha }\left(
\lambda _{\rho }\rho \right) .  \label{Psi}
\end{equation}%
The function $Z(z)$ is normalized by the condition (\ref{Znorm}). In the
special case $D=3$ the subspace with the coordinates $\mathbf{X}$ is absent
and the line element (\ref{ds2cs}) with $b(z)=1$ describes an idealized
cosmic string with zero thickness core in background of Minkowski spacetime.

\subsection{Global monopole type defects}

For global monopole type defects we have $\alpha _{i}=\alpha $, $i=1,\ldots
,D-2$, and in (\ref{ds2}) the subspace $(r,\theta _{1},\ldots ,\theta
_{D-2}) $ is spherically symmetric. In this special case $R_{\theta }=0$ and
the equation for the function $Y_{\mathbf{n}}(\theta )$ coincides with the
corresponding equation for the standard hyperspherical harmonics of degree $%
l $. Denoting these harmonics by $Y(\mathbf{n};\theta )$, for the
corresponding set of quantum numbers we have $\mathbf{n}=(n_{1}\equiv
l,n_{2},\ldots ,n_{D-2})$, with $l=0,1,2,\ldots $, and $-n_{D-3}\leq
n_{D-2}\leq n_{D-3}$. Here, $n_{2},n_{3},\ldots ,n_{D-3}$ are nonnegative
integers arranged in accordance with $n_{i}\leq n_{i-1}\leq l$, $i=3,\ldots
,D-3$. The hyperspherical harmonics satisfy the equation 
\begin{equation}
\Delta _{\theta }Y(\mathbf{n};\theta )=-l(l+D-3)Y(\mathbf{n};\theta ),
\label{EqY}
\end{equation}%
where $\Delta _{\theta }$ is the standard angular Laplacian given by the
angular operator in (\ref{Yeq}) with $\alpha _{i}=1$. Comparing (\ref{Yeq})
and (\ref{EqY}), we conclude that $\gamma =l(l+D-3)/\alpha ^{2}$ in the
special case at hand.

By taking into account that $\int d^{D-2}\theta \,\left\vert Y(\mathbf{n}%
;\theta )\right\vert ^{2}=N(\mathbf{n})$ (for definition of $N(\mathbf{n})$
see \cite{Erde53}), we get the relation 
\begin{equation}
Y_{\mathbf{n}}(\theta )=\alpha ^{1-\frac{D}{2}}\frac{Y(\mathbf{n};\theta )}{%
\sqrt{N(\mathbf{n})}}  \label{Anggm}
\end{equation}%
for the angular part in the mode functions. In the special case $p(r)=r$,
the radial function is given by (\ref{Uflat}) with%
\begin{equation}
\;\nu _{r}=\frac{1}{\alpha }\sqrt{\left( l+\frac{D-3}{2}\right) ^{2}+\left(
D-2\right) \left( D-3\right) \left( \xi -\xi _{D-2}\right) \left( 1-\alpha
^{2}\right) }.  \label{nurgm}
\end{equation}

A composite topological defect combining the cosmic string and global
monopole type defect structures has been discussed in \cite{Beze06,Beze08}.
The corresponding generalization for the line element in warped geometries
becomes%
\begin{equation}
ds^{2}=b^{2}\left( z\right) \left[ -dt^{2}+\sum_{i=1}^{D-4-N}(dX^{i})^{2}+d%
\rho ^{2}+\alpha _{1}^{2}\rho ^{2}d\phi ^{2}+dr^{2}+\alpha
_{2}^{2}r^{2}d\Omega _{N}^{2}+dz^{2}\right] ,  \label{ds2Comp}
\end{equation}%
where $d\Omega _{N}^{2}$ is given by (\ref{dOm}) with the replacements $%
D\rightarrow N+2$ and $\alpha _{i}\rightarrow 1$. This line element combines
two types of defects: a cosmic string with planar angle deficit $2\pi
(1-\alpha _{1})$ in the subspace $(\rho ,\phi )$, and a global monopole in
the subspace $(r,\theta _{1},\ldots ,\theta _{N})$ with solid angle deficit
determined by the parameter $\alpha _{2}$. The part of the scalar field mode
functions in the subspace covered by the coordinates $x^{i}$, $i=0,1,\ldots
,D-1$, is given in \cite{Beze06}. The complete set of mode functions are
obtained by adding the additional factor $Z(z)$ obeying the equation (\ref%
{Zeq}) and normalized by the condition (\ref{Znorm}).

A simpler geometry with combined topological defects is described by 
\begin{equation}
ds^{2}=b^{2}\left( z\right) \left[ -dt^{2}+dr^{2}+\alpha _{1}^{2}r^{2}\left(
d\theta _{1}^{2}+\beta ^{2}\sin ^{2}\theta _{1}d\theta _{2}^{2}\right)
+dz^{2}\right] .  \label{ds2Comp2}
\end{equation}%
This corresponds to a special case of the general metric (\ref{ds2conf})
with $D=4$, $\alpha _{2}=\alpha _{1}\beta $. The scalar field states in $%
(3+1)$-dimensional spacetime, with the line element obtained from (\ref%
{ds2Comp2}) taking $z=\mathrm{const}$ and $b\left( z\right) =1$, have been
recently discussed in \cite{Barb25}.

\subsection{AdS type warp factor}

Consider a warp factor of the form $b(z)=w/z$ with a constant $w$ and $0\leq
z<\infty $. In the absence of defects and for the function $p(r)=r$, this
corresponds to the patch of the AdS spacetime with the curvature radius $w$,
covered by the Poincar\'{e} coordinates (see (\ref{ds2flat})). The
hypersurfaces $z=0$ and $z=\infty $ present the AdS boundary and horizon.
AdS spacetime is the maximally symmetric solution of the Einstein equations
with a negative cosmological constant $\Lambda $ as the only source of the
gravitational field. The cosmological constant and the curvature radius are
related by $\Lambda =-D(D-1)/(2w^{2})$, and the Ricci scalar is expressed as 
$R=-D(D+1)/w^{2}$. The interest in AdS spacetime in gravitational physics
and field theory is motivated by a number of reasons. Due to the high
symmetry of this manifold, a large number of problems in field theory are
exactly solvable on its background. The AdS spacetime also appears as a
ground state in supergravity and string theories. It plays a key role in two
exciting developments of modern theoretical physics, such as braneworld
models and AdS/Conformal field theory (CFT) correspondence. The AdS/CFT
correspondence is a realization of the holographic principle (for reviews
see \cite{Ahar00,Nast15,Ammo15}). It establishes a duality between string or
field theory in AdS bulk and a conformal field theory localized on the AdS
boundary at $z=0$ and provides a powerful tool for the investigation of
non-perturbative effects in fundamental field theories and condensed matter
physics.

Cosmic string and global monopole type topological defects in AdS spacetime
have been discussed in \cite{Dehg02}-\cite{Alva18} (on conical defects in
the context of AdS/CFT correspondence see, e.g., \cite{Aref16,Cres17,Bere23}
and references therein). The polarization of quantum vacuum induced by
cosmic string and global monopole type topological defects in background of
AdS spacetime were studied in \cite{Beze12}-\cite{Oliv24c}. The topological
defect in AdS spacetime will be mapped on a lower dimensional topological
defect in the dual theory. In particular, for a cosmic string in the model
with the line element (\ref{ds2cs}) in 3-dimensional space we will have a
point-like defect in dual theory with spatial dimension $D=2$.

For the function $R_{z}(z)$ in equation (\ref{Zeq}) we get $%
R_{z}(z)=-D(D+1)/z^{2}$ and the solution of equation (\ref{Zeq}), square
integrable at $z=0$ is expressed as%
\begin{equation}
Z(z)=\sqrt{\frac{\lambda _{z}}{2E}}\frac{z^{\frac{D}{2}}}{w^{\frac{D-1}{2}}}%
J_{\nu _{z}}(\lambda _{z}z),\;\nu _{z}=\sqrt{\frac{D^{2}}{4}-D\left(
D+1\right) \xi +w^{2}m^{2}},  \label{Zads}
\end{equation}%
with $0\leq \lambda _{z}<\infty $. The normalization coefficient is
determined from the condition (\ref{Znorm}).

The setup described above can be used to study the influence of topological
defects on the properties of quantum vacuum in braneworld models on the AdS
bulk. In models of the Randall-Sundrum type, branes parallel to the AdS
boundary are present. They impose boundary conditions on bulk quantum
fields. As an example, let us consider a model with a single brane located
at $z=z_{0}$ and with the Robin boundary condition 
\begin{equation}
\left( 1+\beta n^{i}\nabla _{i}\right) \varphi =0,\;z=z_{0},  \label{Rbc}
\end{equation}%
for a bulk quantum scalar field. In (\ref{Rbc}), $n^{i}$ is the inward
pointing normal to the brane and $\beta $ is a constant with dimension of
length. The special cases $\beta =0$ and $\beta \rightarrow \infty $
correspond to Dirichlet and Neumann boundary conditions, respectively.

The brane divides the space into two regions corresponding to $0\leq z\leq
z_{0}$ (region I) and $z_{0}\leq z<\infty $ (region II). The normal in (\ref%
{Rbc}) is given by $n^{i}=\mp \delta _{D}^{i}z_{0}/w$, where the upper and
lower signs correspond to the regions I and II, respectively. In the region
I, the function $Z(z)$ for the integrable scalar modes is still given by (%
\ref{Zads}). The boundary condition (\ref{Rbc}) imposes additional
constraint on the eigenvalues of the quantum number $\lambda _{z}$. They are
roots of the equation%
\begin{equation}
\bar{J}_{\nu _{z}}^{(-)}(\lambda _{z}z_{0})=0.  \label{Jroot}
\end{equation}%
Here and below, for a given function $F(x)$, we use the notation $\bar{F}(x)$
defined by%
\begin{equation}
\bar{F}^{(\mp )}(x)=xF^{\prime }(x)+\beta _{\mp }F(x),\;\beta _{\mp }=\frac{D%
}{2}\mp \frac{w}{\beta },  \label{Fbar}
\end{equation}%
with upper and lower signs corresponding to the regions I and II,
respectively. For simplicity we will assume the values of the dimensionless
ratio $\beta /w$ for which all the roots of (\ref{Jroot}) are real. Let $%
\lambda _{n}$, $n=1,2,\ldots $, be the $n$-th positive root of the equation (%
\ref{Jroot}) with respect to $\lambda _{z}z_{0}$, arranged in the order $%
\lambda _{n}<\lambda _{n+1}$. The integration in the condition (\ref{Znorm})
goes over the region $[0,z_{0}]$ and the normalized function is expressed as%
\begin{equation}
Z(z)=\frac{z^{\frac{D}{2}}J_{\nu _{z}}(\lambda _{n}z/z_{0})}{w^{\frac{D-1}{2}%
}z_{0}\sqrt{E}}\left[ J_{\nu _{z}}^{\prime 2}(\lambda _{n})+\left( 1-\frac{%
\nu _{z}^{2}}{\lambda _{n}^{2}}\right) J_{\nu _{z}}^{2}(\lambda _{n})\right]
^{-\frac{1}{2}},  \label{ZregI}
\end{equation}%
where $E=\sqrt{\lambda _{n}^{2}/z_{0}^{2}+\lambda _{r}^{2}}$. Note that the
derivative $J_{\nu _{z}}^{\prime }(\lambda _{n})$ is expressed in terms of $%
J_{\nu _{z}}(\lambda _{n})$ by using the boundary condition (\ref{Jroot}).

In the region II, the function $Z(z)$ is expressed in terms of the linear
combination of the Bessel function $J_{\nu _{z}}(\lambda _{z}z)$ and the
Neumann function $Y_{\nu _{z}}(\lambda _{z}z)$. The relative coefficient in
the linear combination is determined from the boundary condition (\ref{Rbc})
and the spectrum of the quantum number $\lambda _{z}$ is continuous with $%
0\leq \lambda _{z}<\infty $. The normalization coefficient is determined
from the condition (\ref{Znorm}), where the integration goes over the region 
$[z_{0},\infty )$ and $\delta _{\lambda _{z}\lambda _{z}^{\prime }}=\delta
\left( \lambda _{z}-\lambda _{z}^{\prime }\right) $ in the right-hand side.
In the evaluation of the normalization integral, we note that the dominant
contribution to the integral comes from large values of $z$ and we can
replace the functions $J_{\nu _{z}}(\lambda _{z}z)$ and $Y_{\nu
_{z}}(\lambda _{z}z)$ by their asymptotics for large arguments. In this way
we can show that the normalized mode functions are given by%
\begin{equation}
Z(z)=\sqrt{\frac{\lambda _{z}}{2E}}\frac{z^{\frac{D}{2}}}{w^{\frac{D-1}{2}}}%
\frac{\bar{Y}_{\nu _{z}}^{(+)}(\lambda _{z}z_{0})J_{\nu _{z}}(\lambda _{z}z)-%
\bar{J}_{\nu _{z}}^{(+)}(\lambda _{z}z_{0})Y_{\nu _{z}}(\lambda _{z}z)}{%
\left[ \bar{J}_{\nu _{z}}^{(+)2}(\lambda _{z}z_{0})+\bar{Y}_{\nu
_{z}}^{(+)2}(\lambda _{z}z_{0})\right] ^{1/2}},  \label{ZregII}
\end{equation}%
with the notation defined by (\ref{Fbar}).

\section{The Hadamard function for a scalar field in the geometry of global
monopole on AdS bulk}

\label{sec:MonAdS}

Two-point functions are among the central objects in quantum field theory.
They describe the correlations of quantum fluctuations of fields at
different spacetime points. They can also be used to investigate the
expectation values of various physical observables \cite{Birr82,Park11}. As
an application of the modes given above, here we consider the Hadamard
two-point function for a scalar field around a global monopole in the AdS
bulk. The zero-temperature Hadamard function is defined as the vacuum
expectation value $G(x,x^{\prime })=\left\langle 0\right\vert \varphi
(x)\varphi ^{\dagger }(x^{\prime })+\varphi ^{\dagger }(x^{\prime })\varphi
(x)\left\vert 0\right\rangle $. Expanding the field operator in terms of the
positive and negative energy mode functions $\varphi _{\sigma }^{(\pm )}(x)$
(see (\ref{phiexp})) and by using the rules for the action of the
annihilation and creation operators on the vacuum state, the mode sum
representation%
\begin{equation}
G(x,x^{\prime })=\sum_{\sigma }\sum_{s=\pm }\varphi _{\sigma
}^{(s)}(x)\varphi _{\sigma }^{(s)\ast }(x^{\prime }),  \label{Grep}
\end{equation}%
is obtained.

For a global monopole in $(D+1)$-dimensional AdS spacetime with $D>3$, the
mode functions read%
\begin{equation}
\varphi _{\sigma }^{(\pm )}(x)=\sqrt{\frac{\lambda _{r}\lambda
_{z}r^{D-3}z^{D}}{2EN(\mathbf{n})w^{D-1}\alpha ^{D-2}}}J_{\nu _{z}}(\lambda
_{z}z)J_{\nu _{r}}(\lambda _{r}r)Y(\mathbf{n};\theta )e^{\mp iEt},
\label{phigm}
\end{equation}%
with $E=\sqrt{\lambda _{r}^{2}+\lambda _{z}^{2}}$. The set of quantum
numbers is specified by $\sigma =(\lambda _{r},\lambda _{z},\mathbf{n})$ and%
\begin{equation}
\sum_{\sigma }=\int_{0}^{\infty }d\lambda _{r}\int_{0}^{\infty }d\lambda
_{z}\sum_{\mathbf{n}}.  \label{Sumint}
\end{equation}%
Substituting the modes in (\ref{Grep}), the summation over $\sum_{\mathbf{n}%
} $ is done by using the addition theorem \cite{Erde53}%
\begin{equation}
\sum_{n_{2},\ldots ,n_{D-2}}\frac{Y(\mathbf{n};\theta )}{N(\mathbf{n})}%
Y^{\ast }(\mathbf{n};\theta ^{\prime })=\frac{2l+D-3}{(D-3)S_{D-1}}C_{l}^{%
\frac{D-3}{2}}\left( \cos \Theta \right) ,  \label{Addit}
\end{equation}%
where $C_{l}^{\frac{D-3}{2}}\left( x\right) $ is the Gegenbauer polynomial, $%
\Theta $ is the angle formed by the directions $\theta $ and $\theta
^{\prime }$, and $S_{p}=2\pi ^{p/2}/\Gamma (p/2)$. The Hadamard function is
presented as%
\begin{align}
G(x,x^{\prime }) =&\frac{\left( rr^{\prime }\right) ^{\frac{D-3}{2}}\left(
zz^{\prime }\right) ^{\frac{D}{2}}}{w^{D-1}\alpha ^{D-2}}\sum_{l=0}^{\infty }%
\frac{2l+D-3}{(D-3)S_{D-1}}C_{l}^{\frac{D-3}{2}}\left( \cos \Theta \right) 
\notag \\
&\times \int_{0}^{\infty }d\lambda _{r}\,\lambda _{r}J_{\nu _{r}}(\lambda
_{r}r)J_{\nu _{r}}(\lambda _{r}r^{\prime })  \notag \\
&\times \int_{0}^{\infty }d\lambda _{z}\frac{\lambda _{z}}{E}J_{\nu
_{z}}(\lambda _{z}z)J_{\nu _{z}}(\lambda _{z}z^{\prime })\cos \left( E\Delta
t\right) ,  \label{G1}
\end{align}%
with $\Delta t=t-t^{\prime }$.

For the further transformation of the expression in the right-hand side of (%
\ref{G1}), we use the integral relation \cite{Beze15}%
\begin{equation}
\frac{\cos \left( E\Delta t\right) }{E}=-\frac{1}{2\sqrt{\pi }}\int_{C}\frac{%
ds}{s^{1/2}}e^{-E^{2}s+\frac{\left( \Delta t\right) ^{2}}{4s}},
\label{repcos}
\end{equation}%
where the anticlockwise oriented contour $C$ in the complex plane $%
s=s^{\prime }+is^{\prime \prime }$ consists of two straight segments $s=\pm
i\epsilon $, for $s^{\prime }>0$ and small $\epsilon >0$, and a semicircle
with radius $\epsilon $ that passes around the origin $s=0$ from the left.
The integral over $C$ can also be written in the form $\int_{C}ds=\int_{c_{%
\epsilon }}ds-2\int_{\epsilon }^{\infty }ds$, where $c_{\epsilon }$ is a
circle of radius $\epsilon $ with the center at $s=0$. With the
representation (\ref{repcos}), the integrals over $\lambda _{r}$ and $%
\lambda _{z}$ are evaluated by the formula \cite{Prud2} 
\begin{equation}
\int_{0}^{\infty }d\lambda \,\lambda J_{\nu }(\lambda y)J_{\nu }(\lambda
y^{\prime })e^{-s\lambda ^{2}}=\frac{1}{2s}e^{-\frac{y^{2}+y^{\prime 2}}{4s}%
}I_{\nu }\left( \frac{yy^{\prime }}{2s}\right) ,  \label{IntJ}
\end{equation}%
where $I_{\nu }(x)$ is the modified Bessel function. With these results, by
using the asymptotic of the function $I_{\nu }(x)$ for large argument
(corresponding to small values of $s$ in (\ref{IntJ})), we can see that the
corresponding integral over $c_{\epsilon }$ tends to zero for $\epsilon
\rightarrow 0$ and $\int_{C}ds=-2\int_{0}^{\infty }ds$. Introducing a new
integration variable $u=1/(2s)$, the final expression for the Hadamard
function becomes%
\begin{align}
G(x,x^{\prime }) =&\frac{\Gamma (\frac{D-1}{2})\left( rr^{\prime }\right) ^{%
\frac{D-3}{2}}\left( zz^{\prime }\right) ^{\frac{D}{2}}}{2^{\frac{1}{2}}\pi
^{\frac{D}{2}}(D-3)w^{D-1}\alpha ^{D-2}}\sum_{l=0}^{\infty }\left( l+\frac{%
D-3}{2}\right) C_{l}^{\frac{D-3}{2}}\left( \cos \Theta \right)  \notag \\
&\times \int_{0}^{\infty }du\,\sqrt{u}e^{-\frac{u}{2}\left( r^{2}+r^{\prime
2}+z^{2}+z^{\prime 2}-\left( \Delta t\right) ^{2}\right) }I_{\nu _{r}}\left(
urr^{\prime }\right) I_{\nu _{z}}\left( uzz^{\prime }\right) .  \label{G2}
\end{align}%
This representation is valid in the range $r^{2}+r^{\prime
2}+z^{2}+z^{\prime 2}>\left( \Delta t\right) ^{2}$.

For $D=3$, the geometry under consideration is reduced to (see also (\ref%
{ds2cs}))%
\begin{equation}
ds^{2}=(w/z)^{2}\left( -dt^{2}+dr^{2}+\alpha ^{2}r^{2}d\phi
^{2}+dz^{2}\right) ,  \label{ds2D3}
\end{equation}%
which corresponds to a cosmic string on the AdS bulk. The corresponding mode
functions are obtained from those given in Section \ref{sec:Special} (see (%
\ref{phics}) and (\ref{Zads})): 
\begin{equation}
\varphi _{\sigma }^{(\pm )}(x)=\frac{z^{\frac{3}{2}}e^{in\phi \mp iEt}}{%
2\left( \pi \alpha \right) ^{\frac{1}{2}}w}\sqrt{\frac{\lambda _{r}\lambda
_{z}}{E}}J_{|n|/\alpha }\left( \lambda _{r}r\right) J_{\nu _{z}}(\lambda
_{z}z),  \label{phiD3}
\end{equation}%
with $\nu _{z}=\sqrt{9/4-12\xi +w^{2}m^{2}}$. With these functions, the mode
sum (\ref{Grep}) becomes%
\begin{align}
G(x,x^{\prime }) =&\frac{\left( zz^{\prime }\right) ^{\frac{3}{2}}}{\pi
\alpha w^{2}}\sum_{n=0}^{\infty ^{\prime }}\cos \left( n\Delta \phi \right)
\int_{0}^{\infty }d\lambda _{r}\,\lambda _{r}J_{n/\alpha }\left( \lambda
_{r}r\right) J_{n/\alpha }\left( \lambda _{r}r^{\prime }\right) .  \notag \\
&\times \int_{0}^{\infty }d\lambda _{z}\frac{\lambda _{z}}{E}J_{\nu
_{z}}(\lambda _{z}z)J_{\nu _{z}}(\lambda _{z}z^{\prime })\cos \left( E\Delta
t\right) ,  \label{GD3}
\end{align}%
where $\Delta \phi =\phi -\phi ^{\prime }$. The further transformation of
this expression is the same as that described above for (\ref{G1}). The
resulting expression reads%
\begin{equation}
G(x,x^{\prime })=\frac{2\left( zz^{\prime }\right) ^{\frac{3}{2}}}{\left(
2\pi \right) ^{\frac{3}{2}}w^{2}\alpha }\sum_{n=0}^{\infty ^{\prime }}\cos
\left( n\Delta \phi \right) \int_{0}^{\infty }du\,\sqrt{u}e^{-\frac{u}{2}%
\left( r^{2}+r^{\prime 2}+z^{2}+z^{\prime 2}-\left( \Delta t\right)
^{2}\right) }I_{n/\alpha }\left( urr^{\prime }\right) I_{\nu _{z}}\left(
uzz^{\prime }\right) .  \label{GD32}
\end{equation}%
The two-point functions presented in this section can be used to evaluate
various local properties of the vacuum state in the presence of defects.
These include the vacuum expectation values of the squared field and the
energy-momentum tensor, as well as the self-energy of particles with scalar
charges.

\section{Conclusion}

\label{sec:Conc}

We have considered warped geometries of $(D+1)$-dimensional spacetime with
topological defects. The defects are characterized by a set of parameters $%
(\alpha _{1},\alpha _{2},\ldots ,\alpha _{D-2})$ describing the angular
deficit in corresponding directions. The corresponding metric is defined by (%
\ref{ds2}), (\ref{dOm}), and the components of the Ricci tensor are given by
(\ref{Ricci}). The presence of defects, in general, breaks the spherical
symmetry of the subspace covered by the angular coordinates $(\theta
_{1},\theta _{2},\ldots ,\theta _{D-2})$. Two special cases, most frequently
discussed in the literature, correspond to cosmic strings with the set $%
(1,\ldots ,1,\alpha _{D-2})$ and to a global monopole with $\alpha
_{i}=\alpha $, $i=1,2,\ldots ,D-2$. The presentation (\ref{ds2conf}) shows
the conformal relation between the geometry under consideration and geometry
with the metric tensor independent of the warp coordinate $z$. This
representation is used in braneworld models of the Randall-Sundrum type. The
gravitational field is specified by two functions, $b(z)$ and $p(r)$.

The presence of topological defects modifies the both local and global
properties of the vacuum state for quantum fields. A key step in the
canonical quantization of fields in a given gravitational background is
knowing a complete set of normalized solutions to the classical field
equation. We have considered these solutions for a massive scalar field with
a general curvature coupling parameter. In accordance with the background
field symmetry, the mode functions are presented in a factorized form (\ref%
{phisep}). The separate parts are solutions of the equations (\ref{Ueq}), (%
\ref{Zeq}), and (\ref{Yeq}) and they are normalized by the conditions (\ref%
{Ynorm}), (\ref{Unorm}), and (\ref{Znorm}). The angular part of the mode
functions does not depend on the functions $b(z)$ and $p(r)$ and is
completely determined by set of parameters $\alpha _{i}$. It is further
separated in terms of the associated Legendre function of the first kind
(see (\ref{Y2})).

Special cases of the general setup are considered in section \ref%
{sec:Special}. The first example corresponds to conformally flat spacetimes
in the absence of defects, which are described by the radial function $%
p(r)=r $. For this function and a general defect structure, the solution to
the radial equation is expressed in terms of the Bessel function and is
given by (\ref{Uflat}). As the next example, we have discussed a generalized
cosmic string. In this case, the local geometry outside the defect core
coincides with that in the absence of defects. Introducing cylindrical
coordinates, the line element takes the form (\ref{ds2cs}) and the mode
functions are separated as (\ref{phics}). The normalized solution to the
radial equation is expressed as (\ref{Psi}). The third special example
describes a global monopole in warped spacetime. In this case the angular
subspace is spherically symmetric and the angular part of the mode functions
is expressed in terms of hyperspherical harmonics.

As an important special case of the warp factor, we have considered the
function $b(z)=w/z$, which corresponds to the AdS spacetime in the absence
of defects. For the general case of defects, the $z$-dependent part of the
mode functions is expressed in the form (\ref{Zads}). Using the same
AdS-type warp factor setup, we also considered a geometry with a brane
parallel to the AdS boundary. On the brane, the scalar field obeys a Robin
boundary condition (\ref{Rbc}). The brane divides the space into two
causally separated regions with different properties of the scalar vacuum.
In the region between the brane and the AdS boundary the eigenvalues of the
quantum number $\lambda _{z}$ are quantized by the boundary condition. These
are the roots of the equation (\ref{Jroot}) and the $z$-dependent normalized
factor in the mode functions is given by (\ref{ZregI}). In the region
between the brane and the horizon the spectrum of $\lambda _{z}$ is
continuous and the mode functions are expressed by the formula (\ref{ZregII}%
).

Given the complete set of modes, the two-point functions for a scalar field
are evaluated by using the corresponding mode-sum representations. As an
application of the mode functions described in the paper, we have derived
the formula (\ref{G2}) for the Hadamard two-point function in the global
monopole geometry on the AdS bulk of spatial dimension $D>3$. In the special
case of $D=3$, the geometry of Section \ref{sec:MonAdS} describes a cosmic
string in AdS spacetime. The corresponding Hadamard function is expressed as
(\ref{GD32}). The two-point functions (\ref{G2}) and (\ref{GD32}) can be
used for the investigation of combined effects of background gravitational
field and topological defects on the vacuum polarization.

\section*{Acknowledgment}

The work was supported by the grant No. 21AG-1C047 of the Higher Education
and Science Committee of the Ministry of Education, Science, Culture and
Sport RA. G.V. Mirzoyan was supported by the grant No. 21AG-1C006 of the
Higher Education and Science Committee of the Ministry of Education,
Science, Culture and Sport RA.

\end{document}